\documentclass[a4paper,fleqn]{cas-sc}
\usepackage{amsfonts,mathtools,amssymb,siunitx,mathrsfs}
\usepackage{graphicx, nicefrac}
\usepackage[numbers,sort&compress]{natbib}
\newcommand{\str}{\mathrm{Str}}
\newcommand{\e}{\mathrm{e}}
\newcommand{\jsn}{\mathrm{sn}}
\newcommand{\jcn}{\mathrm{cn}}
\newcommand{\mi}{\mathrm{i}\,}
\usepackage{braket}
\def\tsc#1{\csdef{#1}{\textsc{\lowercase{#1}}\xspace}}
\tsc{WGM}
\tsc{QE}
\tsc{EP}
\tsc{PMS}
\tsc{BEC}
\tsc{DE}

\begin{document}
\let\WriteBookmarks\relax
\def\floatpagepagefraction{1}
\def\textpagefraction{.001}
\shorttitle{The Lax pair for the fermionic Bazhanov-Stroganov $R$-operator}
\shortauthors{A. Melikyan, G. Weber.}

\title [mode = title]{The Lax pair for the fermionic Bazhanov-Stroganov $R$-operator}
                      
\author[1]{\color{black}A. Melikyan}
\cormark[1]
\fnmark[1]
\ead{amelik@gmail.com}
\address[1]{Instituto de F\'isica, Universidade de Bras\'ilia, 70910-900, Bras\'ilia, DF, Brasil}

\author[2]{\color{black}G. Weber}
\cormark[2]
\fnmark[2]
\address[2]{Escola de Engenharia de Lorena, Universidade de S\~ao Paulo, 12602-810, Lorena, SP, Brazil}
\ead{gabrielleweber@usp.br}

\begin{abstract}
We derive the Lax connection of the free fermion model on a lattice starting from the fermionic formulation of Bazhanov-Stroganov's three-parameter elliptic parametrization for the R-operator. It results in the Yang-Baxter and decorated Yang-Baxter equations of difference type in one of the spectral parameters, which is the most suitable form to obtain any relativistic model of free fermions in the continuous limit.
\end{abstract}

\maketitle

\section{Introduction}
One of the remarkable outcomes of the $AdS/CFT$ correspondence (see \cite{Beisert:2010jr} for a review) is the  relation between the $S$-matrix of a spin chain on the gauge theory to the $R$-matrix of the one-dimensional Hubbard 
model \cite{Essler:2005bk,Beisert2007,Rej2006,Martins2007,Mitev2017}. On the other hand, the classical result of Shastry \cite{Shastry1988} is that the construction of the $R$-matrix of the one-dimensional Hubbard 
model involves only the $R$-matrix of the free fermion model. Thus, the integrability of fermionic two-dimensional models -- an interesting subject on its own, becomes especially relevant in relation to the $AdS/CFT$ correspondence. A particularly interesting integrable two-dimensional relativistic purely fermionic model had already appeared before on the string theory side as a result of the fermionization \cite{Arutyunov:2004yx}. Although its various classical and quantum integrability properties have been investigated from various points of view \cite{Melikyan:2011uf,Melikyan:2012kj,Melikyan:2014yma,Melikyan2019ac,Melikyan:2016gkd,Melikyan:2014mfa}, the main challenge still lies in the quantization of the model, due the non-ultralocal nature of the algebra of Lax operators \cite{Melikyan:2012kj,Melikyan:2014yma}. While quantization of such non-ultralocal relativistic fermionic models by means of the standard methods of the integrable systems remains an open problem,  their essential features can already be captured by considering the free fermion model. The explicit expressions for the Lax operators for both full and free models can be found in \cite{Melikyan:2012kj,Melikyan:2014yma}. 

It is not currently known how to formulate non-ultralocal integrable models on a lattice and solve the problem by means of the Bethe Ansatz. The goal of this paper is to address the inverse problem: starting with a suitable known lattice formulation of an integrable model one can simply take the continuous limit and trace the appearance of the non-ultralocal terms in the algebra of the Lax operators. This program can be implemented in particular for the free fermion model, since its lattice formulation is well-know. It becomes especially relevant having in mind the relation outlined above between the $S$-matrix of a spin chain on the gauge theory side and the $R$-matrix of the one-dimensional Hubbard model, which itself reduces to finding the $R$-matrix for the free fermion model (see \cite{Essler:2005bk} for a detailed exposition). However, the resulting $R$-matrix for the one-dimensional Hubbard model is not, unlike most representations of the Yang-Baxter equation, of the difference type in spectral parameters, as a result of the so-called decorated Yang-Baxter equation \cite{Shastry1988}. Thus, it is not obvious how, in principle, to obtain in the continuous limit, which one has to consider in the context of the $AdS/CFT$ correspondence, the $(1+1)$-relativistic fermion model as, for example, the one mentioned above - appearing from string theory, where the dependence of the physical quantities, such as the $S$-matrix is of difference form.

To address this problem, we consider in this paper a more general three-parameter parametrization of the free fermion model due to Bazhanov and Stroganov \cite{Bazhanov:1984iw,Bazhanov:1984ji,Bazhanov:1984jg}. In addition, towards the goal of obtaining a purely fermionic model, we use a more convenient for this purposes fermionic $R$-operator formalism given in \cite{Umeno1998b,Umeno1998,Umeno2000}. The resulting Yang-Baxter and decorated  Yang-Baxter relations turn out to be in the form where the dependence of $R$-matrix is indeed of the difference type with respect to one of the spectral parameters \cite{Melikyan2020}. We then find the Lax connection with the desired dependence of the difference type and bosonize the auxiliary space to obtain the Lax connection in the more familiar graded form, suitable for taking the continuous limit.

\section{Bazhanov-Stroganov elliptic parametrization for the free fermion model}

The free fermion model is defined by the $R$-matrix of an inhomogeneous eight-vertex model of the form:
\begin{align}
    \hat{R}=\begin{pmatrix}
a & 0 & 0 & d\\
0 & b & c & 0 \\
0 & c' & b' & 0 \\
d' & 0 & 0 & a \label{bs:R_matrix_orig},
\end{pmatrix},
\end{align}
together with the following free fermion condition \cite{Fanwu723}:
\begin{align}
    a a' + b b' - c c' - d d'=0 \label{bs:free_fermion_condition}
\end{align}

A particularly interesting and general parametrization of this model has been given by Bazhanov and Stroganov \cite{Bazhanov:1984iw,Bazhanov:1984ji,Bazhanov:1984jg}, where the coefficients in \eqref{bs:R_matrix_orig} are parameterized by the spectral parameter $u \in \mathbb{C}$,  and, in addition, two complex rapidities $\zeta_{1}$ and $\zeta_{2}$:\footnote{Our notations here follow \cite{Melikyan2020}.}
\begin{align}
    &a(u;\zeta_{1},\zeta_{2}) =\rho \left[ 1-\e(u)\e(\zeta_{1})\e(\zeta_{2}) \right], \quad a'(u;\zeta_{1},\zeta_{2}) =\rho \left[ \e(u)-\e(\zeta_{1})\e(\zeta_{2}) \right], \label{bs:a_a_prime} \\
    &b(u;\zeta_{1},\zeta_{2}) =\rho \left[ \e(\zeta_{1})-\e(u)\e(\zeta_{2}) \right], \quad b'(u;\zeta_{1},\zeta_{2}) =\rho \left[ \e(\zeta_{2})-\e(u)\e(\zeta_{1}) \right], \label{bs:b_b_prime}\\
    &c(u;\zeta_{1},\zeta_{2})=c'(u;\zeta_{1},\zeta_{2}) =\rho \: \jsn^{-1}\left(\frac{u}{2}\right)\left[ 1-\e(u)\right]\left[\e(\zeta_{1})e(\zeta_{2})\jsn(\zeta_{1})\jsn(\zeta_{2}) \right]^{1/2}, \label{bs:c_c_prime}\\
    &d(u;\zeta_{1},\zeta_{2})=d'(u;\zeta_{1},\zeta_{2}) =- \mi k \rho \: \jsn \left(\frac{u}{2} \right)\left[ 1+\e(u)\right]\left[\e(\zeta_{1})e(\zeta_{2})\jsn(\zeta_{1})\jsn(\zeta_{2}) \right]^{1/2} \label{bs:d_d_prime}.
\end{align}
Here, the functions $\jsn(x)$ and $\jcn(x)$ are the Jacobi elliptic functions of modulus $\kappa$ \cite{whittaker_watson_1996}, $\e(x):=\jcn(x) + \mi \jsn(x)$ is the elliptic exponential, and $\rho$ is an arbitrary factor. With respect to this parametrization, the $R$-matrix  \eqref{bs:R_matrix_orig} satisfies the Yang-Baxter equation:
\begin{align}
   \hat{R}_{12}(\eta_{12};\zeta_{1},\zeta_{2})\hat{R}_{13}(\eta_{13};\zeta_{1},\zeta_{3})\hat{R}_{23}(\eta_{23};\zeta_{2},\zeta_{3})=\hat{R}_{23}(\eta_{23};\zeta_{2},\zeta_{3})\hat{R}_{13}(\eta_{13};\zeta_{1},\zeta_{3})\hat{R}_{12}(\eta_{12};\zeta_{1},\zeta_{2}),\label{bs:YBE}
\end{align}
which is of difference type with respect to the spectral parameter $u$. In \eqref{bs:YBE}, we have used the shorthand notation $\eta_{jk} \equiv u_{j}-u_{k}$.

In order to obtain a purely fermionic model from the Yang-Baxter equation \eqref{bs:YBE},  it is convenient to introduce from the beginning an equivalent fermionic $R$-operator \cite{Umeno1998b,Umeno1998}, corresponding to the $R$-matrix \eqref{bs:R_matrix_orig}. To this end, one has to apply the Jordan-Wigner transformation (see \cite{Essler:2005bk} for an extensive treatment) to the above $R$-matrix, as well as the the Yang-Baxter equation \eqref{bs:YBE}. The essential technical details are explained in \cite{Umeno1998b,Umeno1998}, and are omitted here. Thus, the fermionic $R$-operator associated to \eqref{bs:R_matrix_orig} takes the form:
\begin{align}
    R_{jk}(u;\zeta_{j},\zeta_{k})&=a(u;\zeta_{j},\zeta_{k})\left[-n_{j} n_{k} \right] +a'(u;\zeta_{j},\zeta_{k})\left[(1-n_{j})(1- n_{k}) \right]+b(u;\zeta_{j},\zeta_{k})\left[n_{j}(1- n_{k}) \right] \nonumber\\
    &+b'(u;\zeta_{j},\zeta_{k})\left[n_{k}(1- n_{j})\right] +c(u;\zeta_{j},\zeta_{k})\left[\Delta_{jk}+\Delta_{kj} \right] + d(u;\zeta_{j},\zeta_{k})\left[-\tilde{\Delta}^{(\dagger)}_{jk}-\tilde{\Delta}_{jk}\right]. \label{bs:fermionic_R}
\end{align}
Here the (spinless) fermionic variables are $c_{k}$, $c^{\dagger}_{k}$ and satisfy the usual anticommutation relations: $\{c_{k},c^{\dagger}_{j}\}=\delta_{jk}$. We have also denoted: $n_{k}=c^{\dagger}_{k}c_{k}$, $\Delta_{jk}=c^{\dagger}_{j}c_{k},\tilde{\Delta}^{(\dagger)}_{jk}=c^{\dagger}_{j}c^{\dagger}_{k}$, and $\tilde{\Delta}_{jk}=c_{j}c_{k}$. Furthermore, it can be shown that the fermionic $R$-operator \eqref{bs:fermionic_R} satisfies the Yang-Baxter equation:
\begin{align}
    R_{12}(\eta_{12};\zeta_{1},\zeta_{2})R_{13}(\eta_{13};\zeta_{1},\zeta_{3})R_{23}(\eta_{23};\zeta_{2},\zeta_{3})=R_{23}(\eta_{23};\zeta_{2},\zeta_{3})R_{13}(\eta_{13};\zeta_{1},\zeta_{3})R_{12}(\eta_{12};\zeta_{1},\zeta_{2}).\label{bs:YBE_fermionic}
\end{align}

Next, we extend the above construction to account for spin degrees of freedom by considering two copies of the $R^{(s)}$-operator, one for each spin $s=\uparrow,\downarrow$, in order to define:
\begin{align}
   \mathcal{R}_{jk}(u_{j}-u_{k};\zeta_{j},\zeta_{k};\zeta'_{j},\zeta'_{k}):=R^{(\uparrow)}_{jk}(u_{j}-u_{k};\zeta_{j},\zeta_{k})R^{(\downarrow)}_{jk}(u_{j}-u_{k};\zeta'_{j},\zeta'_{k}). \label{bs:fermionic_R_spin}
\end{align}
In \eqref{bs:fermionic_R_spin} and following formulas, the parameters with prime stand for the relevant quantities with spin $s=\downarrow$. The fermionic operator $\mathcal{R}_{jk}(u_{j}-u_{k};\zeta_{j},\zeta_{k};\zeta'_{j},\zeta'_{k})$ defined in \eqref{bs:fermionic_R_spin} satisfies the same Yang-Baxter equation \eqref{bs:YBE_fermionic}, and one can construct all relevant quantities following the standard methods.\footnote{See \cite{Umeno1998b} for details on the fermionic $R$-operator corresponding to the $XYZ$ model, and the construction of relevant quantities.} 

As an application, we use the fermionic Yang-Baxter relation for the $R$-operator   \eqref{bs:fermionic_R_spin} with $\zeta_{j} = \zeta_{k} \equiv \zeta$ and $\zeta'_{j} = \zeta_{k}' \equiv \zeta'$ to obtain the Hamiltonian:
\begin{align}
    \hat{\mathcal{H}}=\tau^{-1}(0;\zeta,\zeta')\frac{d}{du}\tau(u;\zeta,\zeta')\vert_{u=0}.\label{bs:Hamiltonian_definition}
\end{align}
The spinful monodromy operator factorizes as: 
\begin{align}
\tau(u;\zeta,\zeta'):=\tau^{(\uparrow)}(u;\zeta) \tau^{(\downarrow)}(u;\zeta'), 
\label{bs:monodromy_op_factorization}
\end{align}
in terms of the monodromy operator $\tau^{(s)}(u;\zeta)$ for spin $s$. Using the explicit form of the coefficients \eqref{bs:a_a_prime}-\eqref{bs:d_d_prime}, and the relations:
\begin{align}
    R^{(s)}_{jk}(0;\zeta)&=\beta(\rho,\zeta) P^{(s)}_{jk},\label{bs:R_betaP}\\
    \tau^{(s)}(0;\zeta)&=\left[\beta(\rho,\zeta)\right]^{N}P^{(s)}_{12}P^{(s)}_{23}\cdot \ldots \cdot P^{(s)}_{N,N-1},\label{bs:monodromy_op}
\end{align}
where we denoted $R^{(s)}_{jk}(0;\zeta):=R^{(s)}_{jk}(0;\zeta,\zeta)$, $\beta(\rho,\zeta):=(-2 \mi \rho) \: \e(\zeta) \: \jsn(\zeta)$, and
\begin{align}
P^{(s)}_{jk}:=1-n_{j,(s)}-n_{k,(s)}+\Delta_{jk,(s)}+\Delta_{kj,(s)}
\label{bs:fermionic_permutation_op} 
\end{align}
is the permutation operator corresponding to spin $s$, one finds from \eqref{bs:Hamiltonian_definition}:\footnote{We consider here periodic boundary conditions identifying the site j=N+1 with the site j=1.}
\begin{align}
\hat{\mathcal{H}}=\frac{1}{\beta(\rho,\zeta)}\sum_{j=1}^{N}\Gamma^{(\uparrow)}_{j,j+1}(\zeta)+\frac{1}{\beta(\rho',\zeta')}\sum_{j=1}^{N}\Gamma^{(\downarrow)}_{j,j+1}(\zeta'),\label{bs:Hamiltonian_Gammas}
\end{align}
with:
\begin{align}
\Gamma^{(s)}_{jk}(\zeta):=P^{(s)}_{jk}\frac{d}{du}R^{(s)}_{jk}(u;\zeta)\Bigr|_{\substack{u=0}}.\label{bs:Gammas_s_def}
\end{align}
The explicit calculation of the functions $\Gamma^{(s)}_{jk}(\zeta)$ in \eqref{bs:Hamiltonian_Gammas} leads to the  Hamiltonian for two non-interacting fermionic $XY$ models in external fields, which are parameterized by the rapidities $\zeta$ and $\zeta'$:\footnote{In passing from \eqref{bs:Hamiltonian_Gammas} to \eqref{bs:XY_Hamiltonian} we have ignored an additive constant factor.}
\begin{align}
   \hat{\mathcal{H}}^{XY}= \sum_{j=1}^{N}\tilde{H}^{(\uparrow)}_{j,j+1}(\zeta)+ \sum_{j=1}^{N}\tilde{H}^{(\downarrow)}_{j,j+1}(\zeta'),\label{bs:XY_Hamiltonian}
\end{align}
where:
\begin{align}
\tilde{H}^{(s)}_{j,j+1}(\zeta)&:=\frac{1}{2\, \jsn(\zeta)}\left[\left(\Delta_{j,j+1,(s)}+\Delta_{j+1,j,(s)}\right)+ k\jsn(\zeta) \left(\tilde{\Delta}^{(\dagger)}_{j,j+1,(s)}-\tilde{\Delta}_{j,j+1,(s)}\right)+2\jcn(\zeta)\left(n_{j,(s)}-\nicefrac{1}{2}\right)\right].\label{bs:XY_Hamiltonian_tilde_s} 
\end{align}

Note that the $\mathcal{R}$-operator \eqref{bs:fermionic_R_spin} is of the difference type in the spectral parameter $u$, unlike the case considered in \cite{Umeno1998b,Umeno1998}. In addition, the procedure in  \cite{Umeno1998b,Umeno1998} to obtain the $XY$ model in an external field \eqref{bs:XY_Hamiltonian_tilde_s} is rather non-linear even for the spinless case, requiring also the decorated Yang-Baxter relation and some nontrivial guess-work, while our construction and derivation of the Hamiltonian \eqref{bs:XY_Hamiltonian} is rather linear and follows the standard steps. We also mention here that the the decorated Yang-Baxter relation considered in \cite{Umeno1998b} is not of the difference type with respect to the spectral parameter $u$, which is the reason for the $R$-matrix of the Hubbard model also not being of the difference type. In contrast, the decorated Yang-Baxter equation corresponding to the fermionic $R$-operator \eqref{bs:fermionic_R} depends on the differences of the spectral parameters $\eta_{jk} \equiv u_{j}-u_{k}$, taking an asymmetrical form only with respect to the other parameters $\zeta_{i}$, and has the following general form \cite{Melikyan2020}:
\begin{align}
    R^{(s)}_{12}(\eta_{12};\zeta_{1},\zeta_{2}-2\mathrm{K}(\kappa);\kappa) \; (2n_{1,s}-1) \; R^{(s)}_{13}(\eta_{13};\zeta_{1},\zeta_{3}-2\mathrm{K}(\kappa);-\kappa) \; R^{(s)}_{23}(\eta_{23};\zeta_{2},\zeta_{3};\kappa) \nonumber \\
    =R^{(s)}_{23}(\eta_{23};\zeta_{2},\zeta_{3};\kappa)\;
    R^{(s)}_{13}(\eta_{13};\zeta_{1},\zeta_{3}-2\mathrm{K}(\kappa);-\kappa)\; (2n_{1,s}-1) \; R^{(s)}_{12}(\eta_{12};\zeta_{1},\zeta_{2}-2\mathrm{K}(\kappa);\kappa), \label{bs:DYBE}
\end{align}
where  we have written the dependence on the modulus $\kappa$ in $R^{(s)}_{jk}(u;\zeta_{j},\zeta_{k};\kappa)$ explicitly, and $\mathrm{K}(\kappa)$ is the complete elliptic integral of the first kind \cite{whittaker_watson_1996}.

\section{The Lax connection}

We now turn to the question of obtaining the Lax connection starting from the Yang-Baxter equation for the fermionic operator $\mathcal{R}_{jk}(u_{j}-u_{k};\zeta_{j},\zeta_{k};\zeta'_{j},\zeta'_{k})$ defined in \eqref{bs:fermionic_R_spin}. We follow the general derivation of the Lax pair outlined in \cite{Izergin1981,Wadati1987,Olmedilla1987,Shiroshi1996}. As in the previous section, we set here $\zeta_{j} = \zeta_{k} \equiv \zeta$ and $\zeta'_{j} = \zeta_{k}' \equiv \zeta'$ to illustrate the main steps, with the general case being a straightforward generalization of the expressions given below. 

Denoting (\rm{cf.} equation \eqref{bs:fermionic_R_spin}):
\begin{align}
\mathcal{R}_{jk}(u;\zeta;\zeta')&:=\mathcal{R}_{jk}(u;\zeta,\zeta;\zeta',\zeta')=R^{(\uparrow)}_{jk}(u;\zeta,\zeta)R^{(\downarrow)}_{jk}(u;\zeta',\zeta'),\label{lax:define_reduced_R}\\
\mathcal{P}_{jk}&:=P^{(\uparrow)}_{jk}P^{(\downarrow)}_{jk},\label{lax:Permutation}
\end{align}
and using the Yang-Baxter equation \eqref{bs:YBE_fermionic} for spin $s=\uparrow, \downarrow$, we find the Yang-Baxter equation for $\mathcal{R}_{jk}(u;\zeta;\zeta')$ in \eqref{lax:define_reduced_R}:
\begin{align}
\mathcal{R}_{12}(u-v;\zeta;\zeta')\mathcal{R}_{13}(u;\zeta;\zeta')\mathcal{R}_{23}(v;\zeta;\zeta')=\mathcal{R}_{23}(v;\zeta;\zeta')\mathcal{R}_{13}(u;\zeta;\zeta')\mathcal{R}_{12}(u-v;\zeta;\zeta').\label{lax:YBE_fermionic}
\end{align}
Then, one finds from \eqref{lax:YBE_fermionic}:
\begin{align}
& \left[ \Gamma_{23}(\zeta;\zeta'),\mathcal{R}_{13}(u;\zeta;\zeta')\mathcal{R}_{12}(u;\zeta;\zeta')\right]\label{lax:commutator_eq} \\
&=\beta(\rho,\zeta)\beta(\rho',\zeta')\left( \frac{d}{dv}\mathcal{R}_{13}(u-v;\zeta;\zeta')\Bigr|_{\substack{v=0}}\mathcal{R}_{12}(u;\zeta;\zeta') -\mathcal{R}_{13}(u-v;\zeta;\zeta')\frac{d}{dv}\mathcal{R}_{12}(u;\zeta;\zeta')\Bigr|_{\substack{v=0}} \right),\notag
\end{align}
where we have denoted (\rm{cf. equation \eqref{bs:Gammas_s_def}}):
\begin{align}
\Gamma_{jk}(\zeta;\zeta'):=\beta(\rho',\zeta')\Gamma^{(\uparrow)}_{jk}(\zeta)+\beta(\rho,\zeta)\Gamma^{(\downarrow)}_{jk}(\zeta').\label{lax:Gamma_def}
\end{align}
From \eqref{bs:Hamiltonian_Gammas} and \eqref{bs:Gammas_s_def} we also find:
\begin{align}
\frac{d}{dt}\mathcal{R}_{jk}(u;\zeta;\zeta') = \frac{\mi}{\beta(\rho,\zeta)\beta(\rho',\zeta')}\left(\left[\Gamma_{j-1,j}(\zeta;\zeta'),\mathcal{R}_{jk}(u;\zeta;\zeta')\right]+\left[\Gamma_{j,j+1}(\zeta;\zeta'),\mathcal{R}_{jk}(u;\zeta;\zeta')\right] \right).\label{lax:Time_derivative}
\end{align}

Finally, renaming the indices $1 \to a;\, 2 \to j;\, 3 \to j+1$ in \eqref{lax:commutator_eq}, and using the equation \eqref{lax:Time_derivative}, we arrive at the zero-curvature condition for integrable models on the lattice: \cite{Faddeev:1987ph,Korepin:1997bk}:
\begin{align}
\frac{d \mathcal{L}_{j}}{dt}= \mathcal{M}_{j+1}\mathcal{L}_{j} - \mathcal{L}_{j}\mathcal{M}_{j},\label{lax:zcc}
\end{align}
where the Lax connection has the following explicit form:
\begin{align}
\mathcal{L}_{j}&=\mathcal{R}_{aj}(u;\zeta;\zeta'),\label{lax:L_operator}\\
\mathcal{M}_{j}&= \frac{\mi}{\beta(\rho,\zeta)\beta(\rho',\zeta')}\mathcal{R}^{-1}_{aj}(u;\zeta;\zeta')\left(\beta(\rho,\zeta)\beta(\rho',\zeta')\frac{d}{dv}\mathcal{R}_{aj}(u-v;\zeta;\zeta')\Bigr|_{\substack{v=0}} - \left[\Gamma_{j-1,j}(\zeta;\zeta'),\mathcal{R}_{aj}(u;\zeta;\zeta') \right] \right).\label{lax:M_operator}
\end{align}
Here $\mathcal{R}^{-1}_{aj}(u;\zeta;\zeta')$ denotes the inverse of \eqref{lax:define_reduced_R}, which corresponds to
\begin{align}
\mathcal{R}^{-1}_{jk}(u;\zeta;\zeta') &= {R^{(\downarrow)}_{jk}}^{-1}(u;\zeta',\zeta'){R^{(\uparrow)}_{jk}}^{-1}(u;\zeta,\zeta), \label{lax:inverse_reduced_R}\\
{R^{(s)}_{jk}}^{-1}(u;\zeta_j,\zeta_k) &= \frac{1}{b b'- c c'} \left[ -a (1 - n_{j,(s)})(1-n_{k,(s)}) + a' n_{j,(s)} n_{k,(s)} + b (1-n_{j,(s)})n_{k,(s)} +b'n_{j,(s)} (1-n_{k,(s)}) \right. \label{lax:inverse_spin_R} \\
&- \left.c \Delta_{jk} -c'\Delta_{kj} + d \tilde{\Delta}^{(\dagger)}_{jk} + d'\tilde{\Delta}_{jk} \right].\nonumber 
\end{align}
For the sake of clarity we omitted in \eqref{lax:inverse_spin_R} the dependence on the spectral parameter $u$ and the rapidities $\zeta_{j}$ and $\zeta_{k}$ of the Boltzmann weights \eqref{bs:a_a_prime} - \eqref{bs:d_d_prime}. We have emphasized the index $a$ in the above formulas, in order to stress that this index corresponds to an extra space, different from the ones corresponding to $j=1, \ldots ,N$.

Thus, we have derived the zero-curvature condition and the corresponding Lax connection \eqref{lax:L_operator} and \eqref{lax:M_operator} starting only from the Yang-Baxter equation \eqref{lax:YBE_fermionic} of the difference type in one of the spectral parameters. We also stress that since the Yang-Baxter \eqref{lax:YBE_fermionic} as well as the decorated Yang-Baxter equations \eqref{bs:DYBE} are of the difference type in one of the spectral parameters, any quantity that is obtained from these two equations (for example, in the context of the Hubbard model to construct an interacting theory) will inherit this dependence.

\section{Jordan-Wigner transformation}
\label{jw}
Up to this point the Lax connection is written entirely in terms of the fermionic operators $c_{j,(s)}$ and $c^{\dagger}_{j,(s)}$. To obtain the usual graded Lax connection in matrix form, we must bosonize the auxiliary space denoted by the index $a$. To this end, we consider the following  Jordan-Wigner transformation \cite{Essler:2005bk}:
\begin{align}
	c_{a,\uparrow} &= c_a \otimes \mathbb{1} \xrightarrow{JW} 
	\bigotimes_{k=1}^{N}\left(-\sigma_k^z\right) \bigotimes_{l=1}^{N}\left(-\tau_l^z\right) \otimes \sigma_a^-, \label{jw:caup} \\
	c_{a,\uparrow}^{\dagger} &= c_a^{\dagger} \otimes \mathbb{1} \xrightarrow{JW} 
	\sigma_a^+ \otimes \bigotimes_{l=1}^{N}\left(-\tau_l^z\right) \bigotimes_{k=1}^{N}\left(-\sigma_k^z\right), \label{jw:caupdg} \\
	c_{a,\downarrow} &= \mathbb{1} \otimes c_a \xrightarrow{JW} 
	(-\sigma_a^z) \otimes \bigotimes_{k=1}^{N}\left(-\sigma_k^z\right) \bigotimes_{l=1}^{N}\left(-\tau_l^z\right) \otimes \tau_a^-, \label{jw:cadown} \\
	c_{a,\downarrow}^{\dagger} &=  \mathbb{1} \otimes c_a^{\dagger}  \xrightarrow{JW} 
	\tau_a^+ \otimes \bigotimes_{l=1}^{N}\left(-\tau_l^z\right) \otimes \left(-\sigma_a^z\right) \otimes \bigotimes_{k=1}^{N}\left(-\sigma_k^z\right),\label{jw:cadowndg}
\end{align}
and apply it only to this auxiliary space to obtain the desired matrix structure. Here $\sigma^i_j$ and $\tau^i_j$, $i=x,y,z$ and $j=a,1,2,\ldots,N$ are two copies of the Pauli matrices, corresponding respectively to spin up and spin down components. Moreover, as usual, we introduce:
\begin{align}
	\sigma^{\pm}_j = \frac{1}{2} \left( \sigma_j^x \pm i \sigma_j^y\right), \; \tau^{\pm}_j = \frac{1}{2} \left( \tau_j^x \pm i \tau_j^y\right).
\end{align} 
We also note that the extra copies of $\sigma_a^z$ appearing in \eqref{jw:cadown} and \eqref{jw:cadowndg} are necessary to ensure the correct anticommutation relations.

Thus, the L-operator \eqref{lax:L_operator} becomes:
\begin{align}
	\mathcal{L}_j = \begin{pmatrix}
		\xi_{j,\uparrow}^{(1)} \xi_{j,\downarrow}^{(1)} & -\Lambda \: \xi_{j,\uparrow}^{(1)} \chi_{j,\downarrow}^{(1)} & \Lambda \: \chi_{j,\uparrow}^{(1)} \xi_{j,\downarrow}^{(1)} & - \chi_{j,\uparrow}^{(1)} \chi_{j,\downarrow}^{(1)} \\
		\Lambda \: \xi_{j,\uparrow}^{(1)} \chi_{j,\downarrow}^{(2)} & \xi_{j,\uparrow}^{(1)} \xi_{j,\downarrow}^{(2)} & \chi_{j,\uparrow}^{(1)} \chi_{j,\downarrow}^{(2)} & \Lambda \: \chi_{j,\uparrow}^{(1)} \xi_{j,\downarrow}^{(2)}\\
		-\Lambda \: \chi_{j,\uparrow}^{(2)} \xi_{j,\downarrow}^{(1)} & - \chi_{j,\uparrow}^{(2)} \chi_{j,\downarrow}^{(1)} & \xi_{j,\uparrow}^{(2)}\xi_{j,\downarrow}^{(1)} & \Lambda \:\xi_{j,\uparrow}^{(2)} \chi_{j,\downarrow}^{(1)} \\
		\chi_{j,\uparrow}^{(2)} \chi_{j,\downarrow}^{(2)} & - \Lambda \: \chi_{j,\uparrow}^{(2)} \xi_{j,\downarrow}^{(2)} & - \Lambda\: \xi_{j,\uparrow}^{(2)} \chi_{j,\downarrow}^{(2)} & \xi_{j,\uparrow}^{(2)}\xi_{j,\downarrow}^{(2)} \label{jw:L-matrix}
	\end{pmatrix}
\end{align}
with 
\begin{align}
\chi_{j,(s)}^{(1)} &= c\left(u;\zeta_{a,(s)},\zeta_{j,(s)} \right) c_{j,(s)} - d\left(u;\zeta_{a,(s)},\zeta_{j,(s)} \right) c^{\dagger}_{j,(s)}, \label{jw:chi_1}\\ 
\chi_{j,(s)}^{(2)} &= d'\left(u;\zeta_{a,(s)},\zeta_{j,(s)} \right) c_{j,(s)} + c'\left(u;\zeta_{a,(s)},\zeta_{j,(s)} \right) c^{\dagger}_{j,(s)}, \label{jw:chi_2}\\ 
\xi_{j,(s)}^{(1)} &= b\left(u;\zeta_{a,(s)},\zeta_{j,(s)} \right) - \left[ a\left(u;\zeta_{a,(s)},\zeta_{j,(s)} \right) + b\left(u;\zeta_{a,(s)},\zeta_{j,(s)} \right) \right]n_{j,(s)}, \label{jw:xi_1}\\
\xi_{j,(s)}^{(2)} &= a'\left(u;\zeta_{a,(s)},\zeta_{j,(s)} \right) + \left[ -a'\left(u;\zeta_{a,(s)},\zeta_{j,(s)} \right) + b'\left(u;\zeta_{a,(s)},\zeta_{j,(s)} \right) \right]n_{j,(s)}, \label{jw:xi_2}
\end{align}
and 
\begin{align}
	\Lambda = \bigotimes_{k=1}^{N} \sigma_k^z \bigotimes_{l=1}^{N} \tau_l^z. \label{jw:lambda}
\end{align}
The factor $\Lambda$ \eqref{jw:lambda} results in the non-local form of  $L$-matrix \eqref{jw:L-matrix}, as it involves contributions from all the sites of the chain. It clearly is a direct consequence of the non-local character of the Jordan-Wigner transformation \eqref{jw:caup} - \eqref{jw:cadowndg}. 

To get rid of this non-locality, we consider the following gauge transformation:
\begin{align}
	\mathcal{L}_j \to G \mathcal{L}_j G^{-1}, \; \text{with} \; G = G_{\uparrow}(\beta_1,\beta_2)\otimes_s G_{\downarrow}(\alpha_1,\alpha_2), \; \alpha_i, \beta_i \in \mathbb{C}, \; i=1,2,  \label{jw:gauge_transformation}
\end{align}
where the gauge transformation acting on each spin component is given by:
\begin{align}
	G_{\uparrow}(\beta_1,\beta_2) = \text{diag}\left(\beta_1 \Lambda, \beta_2\right), \;
	G_{\downarrow}(\alpha_1,\alpha_2) = \text{diag} \left(\alpha_1, \alpha_2 \Lambda \right). \label{jw:gauge_transformation_spin}
\end{align}
The gauge transformed $L$-matrix is local and can be written in terms of the following supertensor product \cite{Essler:2005bk}:
\begin{align}
\mathcal{L}_j =  L_j^{(\uparrow)}(\beta_1,\beta_2) \otimes_s \tilde{L}_j^{(\downarrow)} (\alpha_1, \alpha_2)  \label{jw:local_L-matrix}
\end{align}
of two copies of the spinless $L$-matrix
\begin{align}
L_j^{(s)}(\alpha_1,\alpha_2) = \begin{pmatrix}
\xi_{j,(s)}^{(1)} & \left(\frac{\alpha_1}{\alpha_2}\right) \chi_{j,(s)}^{(1)}\\
\left(\frac{\alpha_2}{\alpha_1}\right) \chi_{j,(s)}^{(2)} & \xi_{j,(s)}^{(2)} 
\end{pmatrix}, \; \text{with} \; \tilde{L}_j^{(s)} (\alpha_1, \alpha_2) = \sigma^z L_j^{(s)}(\alpha_1,\alpha_2) \sigma^z. \label{jw:spinless_L-matrix}
\end{align}
The spinless $L$-matrix \eqref{jw:spinless_L-matrix} can be derived by applying a  spinless version of the Jordan-Wigner transformation defined by \eqref{jw:caup} and \eqref{jw:caupdg} to the spinless $R$-matrix \eqref{bs:fermionic_R} followed by a gauge transformation similar to $G_{\uparrow}(\alpha_1,\alpha_2)$ or $G_{\downarrow}(\alpha_1,\alpha_2)$. It also corresponds to the graded $L$-matrix derived within the formalism of \cite{Essler:2005bk} in terms of graded projection operators.

Before elaborating on this connection, we derive the graded $M$-operator in matrix form. Using the fact that the gauge transformed $L$-operator \eqref{jw:local_L-matrix} factors into the supertensor product of gauge transformed spinless $L$-matrices \eqref{jw:spinless_L-matrix}, the zero curvature condition \eqref{lax:zcc} fixes the form of the $M$-operator as:
\begin{align}
\mathcal{M}_j = \left( M_j^{(\uparrow)}(\beta_1,\beta_2) + \partial_t G_{\uparrow}(\beta_1,\beta_2) G_{\uparrow}^{-1}(\beta_1,\beta_2) \right) \otimes_s \mathbb{1} + \mathbb{1} \otimes_s \left( \tilde{M}_j^{(\downarrow)}(\alpha_1, \alpha_2) + \partial_t G_{\downarrow}(\alpha_1, \alpha_2) G_{\downarrow}^{-1}(\alpha_1, \alpha_2) \right). \label{jw:M-matrix}
\end{align}
Here, the spinless $M$-matrix 
\begin{align}
M_j^{(s)}(\alpha_1,\alpha_2) = \begin{pmatrix}
M^{(s)}_{11,j} & \left(\frac{\alpha_1}{\alpha_2}\right)	M^{(s)}_{12,j}  \\ 
\left(\frac{\alpha_2}{\alpha_1}\right)	M^{(s)}_{21,j} & M^{(s)}_{22,j}
\end{pmatrix}, \; \text{with} \;  \tilde{M}_j^{(s)} (\alpha_1, \alpha_2) = \sigma^z L_j^{(s)}(\alpha_1,\alpha_2) \sigma^z \label{jw:spinless_M-matrix}
\end{align}
can similarly be derived by applying the Jordan-Wigner transformation \eqref{jw:caup} and \eqref{jw:caupdg} followed by the gauge transformation $G_{\uparrow}$ or $G_{\downarrow}$ to the spinless version of the M-operator \eqref{lax:M_operator}. The components of \eqref{jw:spinless_M-matrix} are:
\begin{align}
	M^{(s)}_{11,j} &= \frac{i}{\beta} \frac{1}{b b' - c c'} \Big\{ \beta (b'\dot{b} - c \dot{c}') - (a_0-c_0)c c' n_{j-1,(s)} + \left[ \beta \left(-a' \dot{a} -b'\dot{b} + c \dot{c}'+ d \dot{d}' \right) \right. \Big. \label{jw:M11}\\
	&+ \Big. \left. (a_0-c_0)(c c'- d d') n_{j-1,(s)} \right] n_{j,(s)} 
	+ \left[ (a b' + b b'-c c') \chi^{(4)}_{j-1,(s)} + c d'  \chi^{(3)}_{j-1,(s)}\right]c_{j,(s)} \Big. \nonumber \\
	&+ \Big.\left[ (a' b - b b'+c c') \chi^{(3)}_{j-1,(s)} + c' d  \chi^{(4)}_{j-1,(s)}\right]c_{j,(s)}^{\dagger}
		\Big\},\nonumber \\
	M^{(s)}_{22,j} &= \frac{i}{\beta} \frac{1}{b b' - c c'} \Big\{ \beta (- a\dot{a}' + d'  \dot{d}) + (a_0-c_0)d d' n_{j-1,(s)} + \left[ \beta \left(a \dot{a}' + b\dot{b}' - c' \dot{c} - d' \dot{d} \right) \right. \Big. \label{jw:M22} \\
	&+ \Big. \left. (a_0-c_0)(c c'- d d') n_{j-1,(s)} \right] n_{j,(s)} 
	+ \left[ (a b' + b b'-c c') \chi^{(4)}_{j-1,(s)} + c d'  \chi^{(3)}_{j-1,(s)}\right]c_{j,(s)} \Big.\nonumber \\
	&+ \Big.\left[ (a' b - b b'+c c') \chi^{(3)}_{j-1,(s)} + c' d  \chi^{(4)}_{j-1,(s)}\right]c_{j,(s)}^{\dagger}
	\Big\}, \nonumber \\
	M^{(s)}_{12,j} &=\frac{i}{\beta} \frac{1}{b b' - c c'} \Big\{ a'c \chi^{(3)}_{j-1,(s)} + b'd \chi^{(4)}_{j-1,(s)} + \left[ \beta (b'\dot{c} - c \dot{b}') - (a_0-c_0) b'c n_{j-1,(s)} \right] c_{j,(s)} \Big. \label{jw:M12} \\
	&+ \Big. \left[ \beta (d \dot{a}' - a' \dot{d}) - (a_0-c_0) a'd n_{j-1,(s)} \right] c_{j,(s)}^{\dagger}
	\Big\}, \nonumber \\
	M^{(s)}_{21,j} &=\frac{i}{\beta} \frac{1}{b b' - c c'} \Big\{ b d' \chi^{(3)}_{j-1,(s)} + a c' \chi^{(4)}_{j-1,(s)} + \left[ \beta (-a\dot{d}' + d' \dot{a}) + (a_0-c_0) a d' n_{j-1,(s)} \right] c_{j,(s)} \Big. \label{jw:M21} \\
	&+ \Big. \left[ \beta (- c' \dot{b} + b \dot{c}') + (a_0-c_0) b c' n_{j-1,(s)} \right] c_{j,(s)}^{\dagger} \Big\}. \nonumber
\end{align}
To avoid cluttering, we omitted all the arguments of the Boltzmann weights as well as of the quantities derived thereof, such as the nonzero coefficients $a_0, b_0, c_0, d_0$ of $\Gamma_{j-1,j}^{(s)}(\zeta)$ as in \eqref{bs:Gammas_s_def} and the coefficients $\dot{a}, \dot{a}', \dot{b}, \dot{b}'$, $\dot{c}, \dot{c}', \dot{d}, \dot{d}'$ of $\partial_{v} R^{(s)}_{a j}(u-v,\zeta_a, \zeta_j)|_{v=0}$. The parameter $\beta$ also depends on $\rho$ and $\zeta$ as in \eqref{lax:M_operator}. We also introduced the quantities:
\begin{align}
	\chi^{(3)}_{j,(s)} &= b_0(\zeta) c_{j,(s)} - d_0(\zeta) c_{j,(s)}^{\dagger},\\
	\chi^{(4)}_{j,(s)} &= - d_0(\zeta) c_{j,(s)} + b_0(\zeta) c_{j,(s)}^{\dagger}.
\end{align}
The contribution from the derivatives of the gauge matrices $G_{\uparrow}$ and $G_{\downarrow}$ to \eqref{jw:M-matrix} can be easily computed as:
\begin{align}
\left( \partial_t G_{\uparrow} G_{\uparrow}^{-1} \right) \otimes_s \mathbb{1} + \mathbb{1} \otimes_s  \left( \partial_t G_{\downarrow} G_{\downarrow}^{-1} \right) = \text{diag} \left( \partial_t \Lambda \: \Lambda, 2 \partial_t \Lambda \: \Lambda, 0, \partial_t \Lambda \:  \Lambda\right),
\end{align}
where
\begin{align}
	\partial_t \Lambda \: \Lambda &= 2 \sum_{j=1}^{N} \sum_{s=\uparrow,\downarrow} \left[ \partial_t n_{j,(s)}, n_{j,(s)} \right] = 4 i \sum_{j=1}^{N} \sum_{s=\uparrow,\downarrow} \left[ b_0 \left(\Delta_{j,j+1,(s)}+\Delta_{j+1,j,(s)}\right) + d_0 \left(\tilde{\Delta}^{(\dagger)}_{j,j+1,(s)}-\tilde{\Delta}_{j,j+1,(s)}\right) \right]
\end{align}
corresponds to a multiple of the Hamiltonian of the $XY$-model.

Finally, to elaborate on the connection with the formalism of \cite{Essler:2005bk}, we consider the invariance of the Yang-Baxter relations  \eqref{bs:YBE} under the simultaneous redefinition of the Boltzmann weights: $a \leftrightarrow -a$, $a' \leftrightarrow -a'$, $c \leftrightarrow -c$, $c' \leftrightarrow -c'$, to define an equivalent representation of \eqref{bs:YBE}. Thus, denoting the $R$-matrix \eqref{bs:R_matrix_orig} elements as:
\begin{align}
\hat{R}^{11}_{11} = -a, \; \hat{R}^{11}_{22} = d, \; \hat{R}^{12}_{12} = b, \; \hat{R}^{12}_{21} = -c, \; \hat{R}^{21}_{12} = - c', \; \hat{R}^{21}_{21} = b', \; \hat{R}^{22}_{11} = d', \; \hat{R}^{22}_{22} = -a', \label{jw:R_matrix_parametrization}
\end{align}
it is easy to verify that it satisfies the compatibility condition of Kulish and Sklyanin \cite{Kulish:1980ii}:
\begin{align}
\hat{R}^{\alpha \beta}_{\gamma \delta}(u;\zeta_j,\zeta_k) = (-1)^{p_{\alpha} + p_{\beta} + p_{\gamma} + p_{\delta}} \hat{R}^{\alpha \beta}_{\gamma \delta}(u;\zeta_j,\zeta_k), \label{jw:compatibility_condition}
\end{align}
which forces some elements of the $R$-matrix to vanish, so that it is compatible with the grading of the underlying vector space. Here, $p$ is a parity function defined on the homogeneous components of a finite dimensional local space of states $V = V_1 \oplus V_2$, so that $p_{\alpha} := p\left(\mathbf{v}_{\alpha}\right), \; \mathbf{v}_\alpha \in V_{\alpha}, \; \alpha=1,2$. For the case under consideration \eqref{jw:R_matrix_parametrization}, we have $p_1=0, \; p_2 =1$. Thus, it is possible to define a graded $L$-matrix at site $j$ as \cite{Essler:2005bk}:
\begin{align}
{L_j}_{\beta}^{\alpha}(u;\zeta_a,\zeta_j) = (-1)^{p_{\alpha} p_{\gamma}} \hat{R}^{\alpha \gamma}_{\beta \delta}(u;\zeta_a,\zeta_j) \: {e_j}^{\delta}_{\gamma}, \label{jw:L-korepin}
\end{align}
so that it satisfies the usual bilinear relations:
\begin{align}
\tilde{R}(\eta_{ab};\zeta_a,\zeta_b) \left( L_j(\eta_{aj};\zeta_a,\zeta_j) \otimes_s L_j (\eta_{bj};\zeta_b,\zeta_j) \right) = \left( L_j(\eta_{bj};\zeta_b,\zeta_j) \otimes_s L_j(\eta_{aj};\zeta_a,\zeta_j)\right) \tilde{R}(\eta_{ab};\zeta_a,\zeta_b), \label{jw:L_bilinear_relations}
\end{align}
where $\tilde{R}^{\alpha \beta}_{\gamma \delta}(u;\zeta_j,\zeta_k) = \hat{R}^{\beta \alpha}_{\gamma \delta}(u;\zeta_j,\zeta_k)$.

The graded projection operators ${e_j}^{\beta}_{\alpha}$ appearing in \eqref{jw:L-korepin} can be defined through the anticommutation relations:
\begin{align}
{e_j}_{\alpha}^{\beta} {e_j}_{\gamma}^{\delta} &= \delta_{\gamma}^{\beta} {e_j}_{\alpha}^{\delta}, \label{jw:graded_projection_operators_algebra}\\
{e_j}_{\alpha}^{\beta} {e_k}_{\gamma}^{\delta} &= (-1)^{\left(p_{\alpha} + p_{\beta}\right) \left(p_{\gamma} + p_{\delta}\right)} {e_k}_{\gamma}^{\delta} {e_j}_{\alpha}^{\beta} \nonumber. 
\end{align}
A possible matrix representation in terms of fermionic creation and annihilation operators of the algebra \eqref{jw:graded_projection_operators_algebra} is 
\begin{align}
e_j = \begin{pmatrix}
c^{\dagger}_{j} c_j & c^{\dagger}_j \\
c_j & 1 - c^{\dagger}_{j} c_j
\end{pmatrix}. \label{jw:graded_projection_operator_rep}
\end{align}
Plugging \eqref{jw:graded_projection_operator_rep} into \eqref{jw:L-korepin} with the parametrization \eqref{jw:R_matrix_parametrization} leads to the spinless $L$-matrix \eqref{jw:spinless_L-matrix} for $\alpha_1 = -1$ and $\alpha_2 = 1$.

To conclude this section, we note that the $L$-matrix \eqref{jw:spinless_L-matrix} reduces to that of the $XY$-model, which is the usual building block for the construction of the $R$-matrix for the Hubbard model. This special case can be obtained by fixing $k=0$ and $\zeta_1 = \zeta_2=\nicefrac{\pi}{2}$, and normalizing the Boltzmann weights with respect to $c(u)$. The resulting Boltzmann weights are:
\begin{align}
a(u) = a'(u) = \cos \frac{u}{2}, \; b(u) = b'(u) = \pm \sin \frac{u}{2}, \; c(u) = 1, \; d(u) =0. \label{jw:XY_limit}
\end{align} 


%

\section{Conclusion}
We have constructed the Lax connection for the free fermion model starting from the fermionic form of Bazhanov and Stroganov's solution for the Yang-Baxter equation, which is of difference type in one of the spectral parameters and is most suitable for obtaining a relativistic theory in the continuous limit. We have employed Umeno's fermionic $R$-matrix formalism as it immediately results in the fermionic form of the Lax connection, thus, making the procedure of taking the continuous limit a rather straightforward calculation using the explicit expressions given in section \ref{jw}. We therefore have made a step forward towards relating the Lax connection of the continuous fermionic model in \cite{Melikyan:2012kj} to a lattice model, with the larger goal, as discussed in the introduction, of quantizing non-ultralocal models. The results of this investigation will be presented in a future publication.

\bibliographystyle{elsarticle-num}

\bibliography{bs_lax}

\end{document}